\documentclass[a4paper,11pt]{article}
\pdfoutput=1 
\usepackage{jheppub} 
\usepackage[T1]{fontenc} 

\title{\boldmath Top quark mass measurements at the $t\bar{t}$ threshold with CEPC}

\author[a,b]{Zhan Li,}
\author[c,1]{Xiaohu Sun,\note{Corresponding author.}}
\author[a,b,1]{Yaquan Fang,}
\author[a,b]{Gang Li,}
\author[a,b]{Shuiting Xin,}
\author[a,b]{Shudong Wang,}
\author[a,b]{Yiwei Wang,}
\author[a,b]{Yuan Zhang,}
\author[a,b]{Hao Zhang,}
\author[a,b]{Zhijun Liang}

\affiliation[a]{Institute of High Energy Physics, 19B, Yuquan Road, Shijing District, Beijing 100049, China}
\affiliation[b]{University of Chinese Academy of Sciences (CAS), 19A Yuquan Road, Shijing District, Beijing 100049, China}
\affiliation[c]{Department of Physics and State Key Laboratory of Nuclear Physics and Technology, Peking University, 209 Chengfu Road, Haidian District, Beijing 100871, China}

\emailAdd{Xiaohu.Sun@pku.edu.cn}
\emailAdd{fangyq@ihep.ac.cn}

\abstract{We present a study of top quark mass measurements at the $t\bar{t}$ threshold based on CEPC. A centre-of-mass energy scan near two times of the top mass is performed and the measurement precision of top quark mass, width and $\alpha_S$ are evaluated using the $t\bar{t}$ production rates. Realistic scan strategies at the threshold are discussed to maximise the sensitivity to the measurement of the top quark properties individually and simultaneously in the CEPC scenarios assuming a limited total luminosity of 100 fb$^{-1}$. With the optimal scan for individual property measurements, the top quark mass precision is expected to be 9~MeV, the top quark width precision is expected to be 26~MeV, and $\alpha_S$ can be measured at a precision of 0.00039. Taking into account the uncertainties from theory, background subtraction, beam energy and luminosity spectrum, the top quark mass can be measured at a precision of 14 MeV optimistically and 34 MeV conservatively at CEPC.}

\begin{document} 
\maketitle
\flushbottom

\section{Introduction}
\label{sec:intro}

Top quark, the heaviest fundamental particle observed so far, plays an important role in the Standard Model (SM). It provides the strongest coupling to the SM Higgs boson and opens doors to new physics beyond the SM (BSM). Till now, the top quark mass have only been measured at hadron collisions, e.g. the Tevatron and the Large Hadron Collider (LHC), using the direct reconstruction of the invariant mass of the top quark decay products. In future electron-positron colliders the top quark mass can be measured not only by the direct reconstruction but also by a scan on the centre-of-mass energy at the \ttbar threshold. The cross-section of \ttbar increases sharply as the centre-of-mass energy going through the \ttbar threshold and depends strongly on the top quark mass, width and $\alpha_S$, which provides a sensitive probe to these measurements. This is the so-called threshold-scan method that was discussed for top quark mass measurements at an electron-positron collider~\cite{Fadin:1987wz,Fadin:1988fn,Strassler:1990nw,Bigi:1986jk}.

In experiments, the top quark mass has been measured by using the direct reconstruction of the top quark decay products as 174.30 $\pm$ 0.35 (stat.) $\pm$ 0.54 (syst.) GeV from the combined results of CDF and D0 at Tevatron~\cite{CDF:2016vzt},
172.69 $\pm$ 0.25 (stat.) $\pm$ 0.41 (syst.) GeV with ATLAS~\cite{ATLAS:2018fwq} and 172.44 $\pm$ 0.13 (stat.) $\pm$ 0.47 (syst.) GeV with CMS~\cite{CMS:2015lbj} at the LHC. The precision till now is about half a GeV and it is mainly limited by the systematic uncertainties that are not easily reduced in the future. On the contrary, the threshold-scan method has been widely used \cite{LEP:2020w, BESIII:2019} and shown good performance with a precision of top quark mass measurement at about 20-30 MeV that was studied previously with ILC, CLIC and FCC-ee~\cite{Martinez:2002st,Seidel:2013sqa,Horiguchi:2013wra,CLICdp:2018esa,FCC:2018evy}.

The threshold-scan method also provides a theoretically well defined mass that can be calculated with a high degree of precision and can be easily converted to various theoretical schemes. This cannot be realised in the reconstructed top mass peak method in which the generated mass peak is usually used as a template to fit to the observed data, since the generator mass is not well-defined theoretically.

In this article, we discuss the threshold-scan method and propose realistic scan strategies for the top quark mass measurements with electron-positron collisions based on the Circular Electron Positron Collider (CEPC). The experimental conditions at CEPC are introduced in Sec.~\ref{sec:cepc}.
The threshold-scan method applied to the CEPC senarios, the realistic scan strategies and the optimal precision in top quark measurements are discussed in Sec.\ref{sec:threshold}.
The systematic uncertainties from the theoretical calculation on the cross-section, the beam energy, the luminosity spectrum and the background contamination are discussed in Sec.~\ref{sec:syst}. Eventually the conclusions are presented in Sec.~\ref{sec:conclu}.

\section{Experimental conditions at CEPC}
\label{sec:cepc}

CEPC is a large concept collider with a circumference of 100 km and two interaction points~\cite{CEPCStudyGroup:2018rmc,CEPCStudyGroup:2018ghi}.
The accelerator complex consists of a linear accelerator (Linac), a damping ring (DR), the booster, the collider and several transport lines.
The center-of-mass energy of CEPC will be 240 GeV, at which collision energy it will serve as a Higgs factory, generating more than one million Higgs particles corresponding to a total luminosity of 5.6 ab$^{-1}$ as a baseline design. The design also allows operation at 91 GeV for a Z factory and at 160 GeV for the $W^+ W^-$ threshold scan. The number of Z particles will be close to 1 trillion from 16 ab$^{-1}$ data taking, and $W^+ W^-$ pairs about 15 million from 2.6 ab$^{-1}$ data taking. These unprecedented large number of particles make the CEPC a powerful instrument not only for precision measurements on these important particles, but also in the search for new physics.
Apart from those, CEPC could also ramp up the center-of-mass energy to reach the $t\bar{t}$ threshold. A preliminary plan is to collect a total luminosity of 100 fb$^{-1}$.

\begin{figure}[tbp]
\centering
\includegraphics[width=0.90\textwidth]{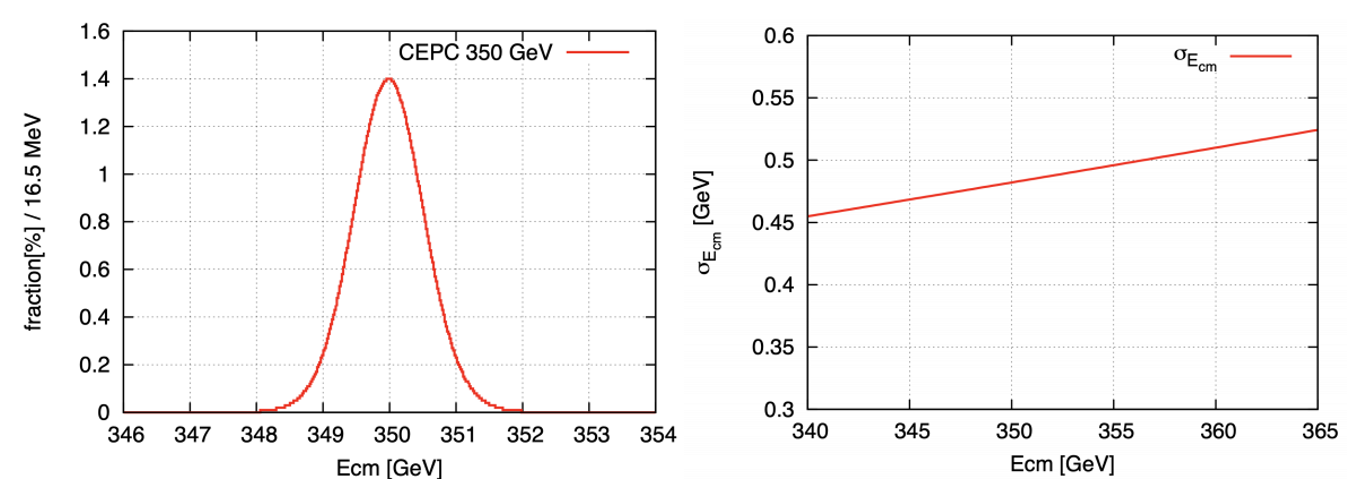}
\caption{\label{fig:ls-cepc} The luminosity spectrum of CEPC, the spread of which increases as the centre-of-mass energy rises.}
\end{figure}

Unlike linear colliders, circular colliders have bending magnets in the acceleration ring, which help to constrain the beam energy spread. Thus, the luminosity spectrum (LS) is more concentrated at the beam energy peak. The spread of LS increases as the centre-of-mass energy rises. At CEPC, LS is modeled by a simple Gaussian distribution and its width $\sigma_{LS}$ is a function of centre-of-mass energy $\sqrt{s}$:
\begin{equation}
\label{eq:sigma_ls}
    \sigma_{LS}=0.51\times (\frac{\sqrt{s}}{360})^{2}
\end{equation}
which is shown in Fig.~\ref{fig:ls-cepc}.
Thus, $\sigma_{LS}$ is roughly 500 MeV around the $t\bar{t}$ threshold.
resulting in a luminosity spectrum with more than 90 \% of the luminosity in the top 1 \% of the energy at 350 GeV, compared to 77 \% of that in the top 1 \% at CLIC, at the same energy~\cite{Linssen:1425915,Aicheler:1500095}.

\section{Top quark mass measurements at the threshold}
\label{sec:threshold}


The package QQbar\_threshold (VERSION 2.2.0) is used~\cite{Beneke:2016kkb,Beneke:2017rdn} to calculate the \ttbar cross section around the threshold. The potential-subtracted (PS) scheme is used for the top quark mass, $m_{top}^{PS} = 171.5$~GeV. The width is set to $1.33$~GeV and the strong coupling $\alpha_S$ to $0.1184$. The cross section is calculated in N$^3$LO QCD in resumed non-relativistic perturbation theory~\cite{Beneke:2013jia,Beneke:2015kwa} and NNLO in electroweak~\cite{Beneke:2017rdn}. Two important beam effects, e.g. the initial state radiation (ISR) and the LS that is a function of centre-of-mass energy based on the CEPC scenario, are both taken into account with QQbar\_threshold. 

\begin{figure}[tbp]
\centering
\includegraphics[width=0.70\textwidth]{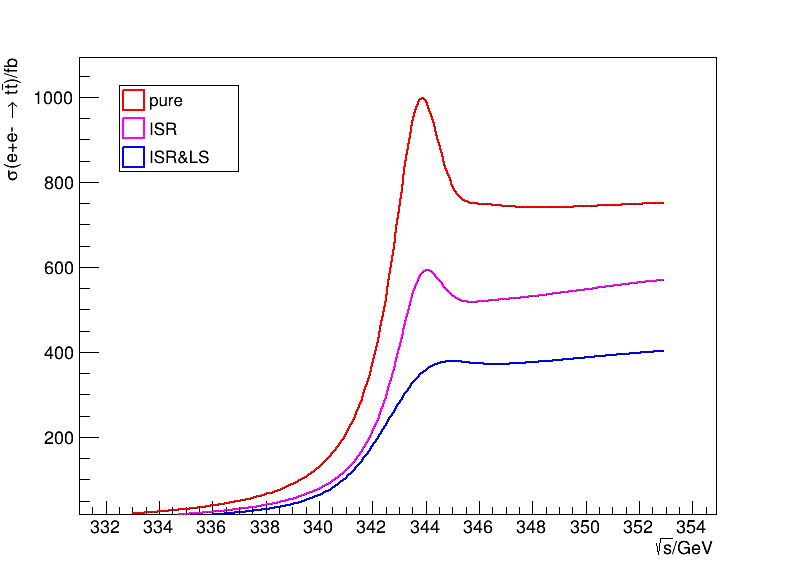}
\caption{\label{fig:ls-xs} The \ttbar cross-section as a function of centre-of-mass energy calculated from QQbar\_threshold including the cross-section values without ISR or LR, the ones with ISR only and the ones with both ISR and LS.}
\end{figure}

The cross-section as a function of centre-of-mass energy is shown in Fig.~\ref{fig:ls-xs} including the calculation without ISR or LS, the one with ISR and the one with both ISR and LS. The original cross-section curve is significantly worn down by the two effects both smearing out the energy distribution of the beam particles.

The number of \ttbar events are calculated with the cross-section that is a function of top quark mass, width and $\alpha_S$, and the corresponding luminosity assumed for the centre-of-mass energy in the scan. The semi-leptonic and full-hadronic decay modes are taken into account. The signal efficiency and acceptance follow the same as evaluated in Ref.\cite{Seidel:2013sqa}. The background contribution is relatively low and can be subtracted as discussed in Ref.\cite{Seidel:2013sqa}. The background is neglected for the nominal estimations in this section but its impact will be discussed as systematic uncertainties later in Sec.~\ref{sec:syst}.


To extract the top quark mass, width and $\alpha_S$, the number of events are counted at each centre-of-mass energy in the scan. A likelihood function is constructed to perform the fits with the counted numbers, between the expected number of events evaluated with the cross-section turn-on curve and the observed number of events from the CEPC. The latter number is simply set equal to the former to make the nominal evaluation and estimate the measurement precision, i.e. the one-sigma error in in the likelihood curve. The likelihood function is defined as

\begin{equation}
\label{eq:lh}
    \mathcal{L} = \prod_{i=1}^{N} P( D | \sigma_{t\bar{t}}(m_{top}, \Gamma_{top}, \alpha_S, \sqrt{s_i}) \times L_i \times \epsilon)
\end{equation}

\noindent
where the observed number of events ($D$) should follow the Poissonian distribution with the expected mean as $E = \sigma_{t\bar{t}}(m_{top}, \Gamma_{top}, \alpha_S, \sqrt{s_i}) \times L \times \epsilon$ under certain centre-of-mass energy ($\sqrt{s_i}$ indexed with $i$). The likelihood function combines all $N$ energy points by multiplying all the Poissonian probability $P$ under each collision energy. In the equation, $\sigma_{t\bar{t}}$ stands for the cross-section, $m_{top}$ for the top quark mass, $\Gamma_{top}$ for the top quark width, $\alpha_S$ for the strong coupling, $L_i$ for the luminosity allocated to the collisions at the centre-of-mass energy of $\sqrt{s_i}$, and $\epsilon$ for the selection efficiency times acceptance of the \ttbar signal events. The precision of measurement on $m_{top}, \Gamma_{top}$, and $\alpha_S$ is evaluated by minimising the negative log likelihood function.


\begin{figure}[tbp!]
\centering
\includegraphics[width=0.70\textwidth]{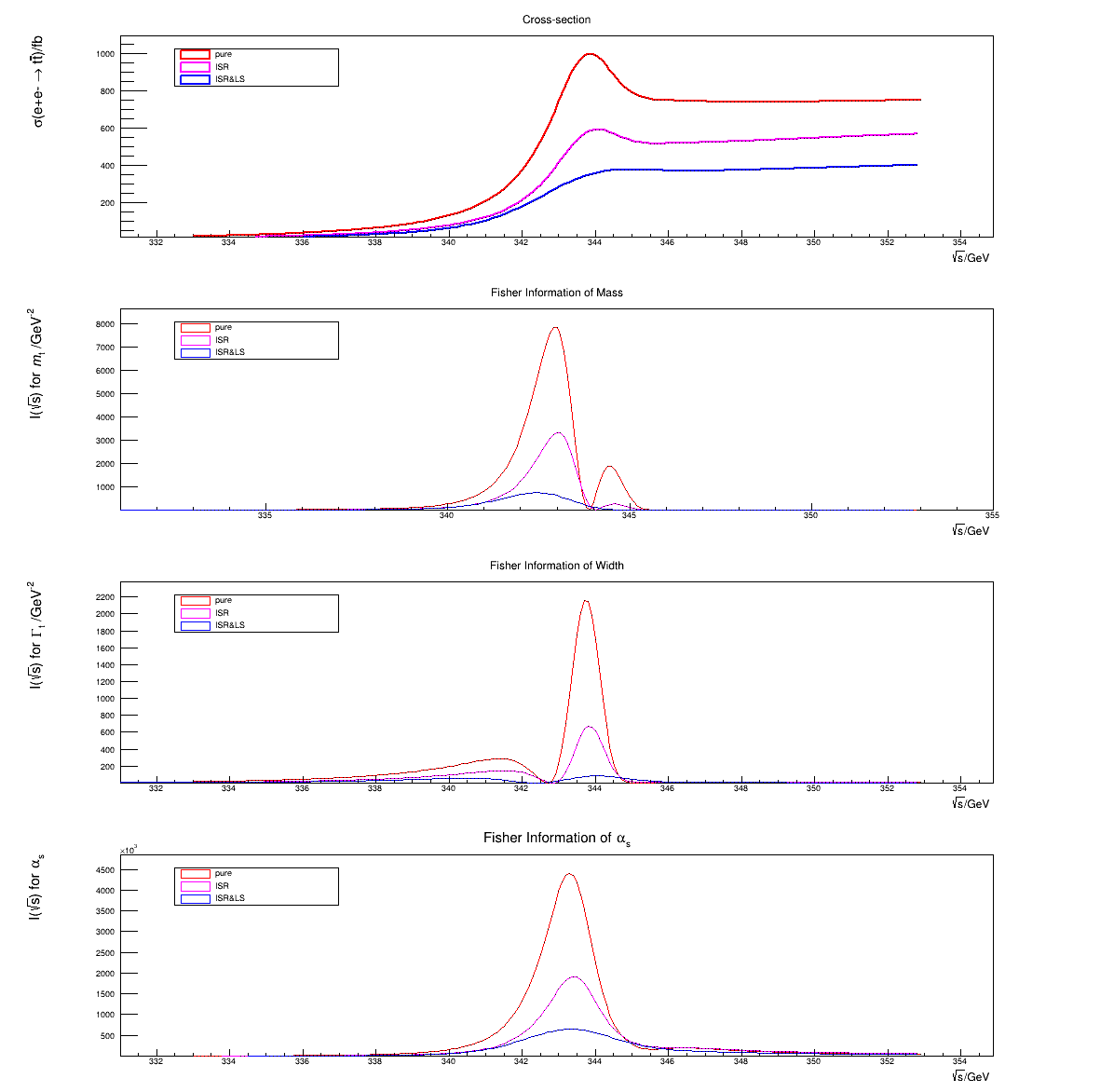}
\caption{\label{fig:fisher} The Fisher information of top-quark mass, width and $\alpha_S$ along the centre-of-mass energy around the \ttbar threshold. }
\end{figure}

The cross-section turn-on curve contains different amounts of information for top quark mass, width and $\alpha_S$ along the centre-of-mass energy scan points. One needs to find out what level of sensitivity the cross-section curve can provide for the measurements.
The Fisher information is used here as a rule of thumb. At a certain centre-of-mass energy ($\sqrt{s}$) one can consider the measured cross-section ($\sigma$) as a random variable which follows a Gaussian distribution ($G$) with its mean value centred at the true cross-section ($\sigma_{0}(\sqrt{s},\theta)$, where $\theta$ can be top quark properties like top quark mass $m_{top}$ and width $\Gamma_{top}$ as well as the strong coupling $\alpha_S$). Thus the Fisher information reads

\begin{equation}
\label{eq:FI}
I(\sqrt{s}) = \int (\frac{\partial log(G(\sigma | \sigma_{0}(\sqrt{s},\theta),\sqrt{\sigma_{0}(\sqrt{s},\theta)}))}{\partial\theta})^{2}\times G(\sigma | \sigma_{0}(\sqrt{s},\theta), \sqrt{\sigma_{0}(\sqrt{s},\theta)})d\sigma
\end{equation}

In this way, the Fisher information can reflect the sensitivity to the measurements of top quark mass, width and $\alpha_S$ as a function of centre-of-mass energy, respectively, as shown in Fig.~\ref{fig:fisher}.
The larger the value in the Y-axis of the Fisher information is, the more sensitive to the relevant measurement the cross-section at this centre-of-mass energy would be. The cross-section curve is found most sensitive to top quark mass when the cross-section ramps up around the threshold, and is sensitive to top quark width when the cross-section peaks, while it is sensitive to $\alpha_S$ by its overall rate thus mainly close to the cross-section peak.

The effects of ISR and LS wear out the original cross-section curve resulting significant drops in Fish information for all three parameters as shown in Fig.\ref{fig:fisher}. 
Considering the ISR effect only, collisions at $\sqrt{s}$ = 343.00 GeV provides the highest sensitivity to the top quark mass, $\sqrt{s}$ = 343.75 GeV for top quark width and $\sqrt{s}$ = 343.5 GeV for $\alpha_S$.
Considering both ISR and LS effects, the optimal energy points are shifted, mostly due to the fact that the spread of LS is not a constant, instead, increases as the energy rises. Then collisions at $\sqrt{s}$ = 342.75 GeV provides the highest sensitivity to the top quark mass measurements, $\sqrt{s}$ = 344.00 GeV for top quark width, and $\sqrt{s}$ = 343.25 GeV for $\alpha_S$.



With the total luminosity limited to 100 pb$^{-1}$, we discuss the optimal scan strategy with only statistical uncertainty in this section. Firstly, the luminosity is evenly allocated to each centre-of-mass energy scan point.
Using the Fisher information as a guide, one can propose various grids of collision energy and evaluate the sensitivities.
The following grids are tested.
\begin{itemize}
\item 8-point grid: \{341, 342, 342.5, 342.75, 343, 343.5, 344.5, 345\} GeV
\item 6-point grid: \{342, 342.5, 342.75, 343, 343.5, 344.5\} GeV
\item 4-point grid: \{342.5, 342.75, 343, 343.5\} GeV
\item 1-point grid: \{342.75\} GeV
\end{itemize}
\noindent
among which the energy point most sensitive to top mass that is indicated by Fisher information is included.
The likelihood function defined in Eq.~\ref{eq:lh} is calculated for each scan grid and the error at 68\% confidence level in the likelihood scan is taken as the statistical uncertainty.
The uncertainties on top mass measurement are then 13 MeV from 8 energy points, 12 MeV from 6 energy points, 10 MeV from 4 energy points and 9 MeV from 1 energy point.
Therefore, when allocating luminosity evenly to the centre-of-mass energy points in the scan, the optimal solution is to take all the data at the energy point that is most sensitive to top mass given that the optimal energy point is known. Further tests are performed for unevenly allocating the luminosity around the optimal energy point, the conclusion does not change.

To find the optimal energy point that has the most sensitivity to the top quark mass measurement, a simple low-luminosity scan can be performed. By the time of CEPC, the LHC or HL-LHC should be able to provide the top quark mass measurement with a certain level of precision. According to that one can scan evenly in the range with very low luminosity and then sort the scan points in the order of the top quark mass measurement error, just like the list in Tab.~\ref{tab:low-lumi-scan}, where 6 energy points, $\sqrt{s}=$~\{340, 341, 342, 343, 344, 345\}, are scanned with even luminosity of 1 fb$^{-1}$ per energy point and the top three are shown.
From this scan, one can identify the optimal energy point should sit between 342 GeV and 343 GeV. Then a finer grid between 342 GeV and 343 GeV, $\sqrt{s}=$~\{342.25, 342.5, 342.75, 343, 343.25, 343.5\}, can be scanned, again with even luminosity of 1 fb$^{-1}$ per energy point, as shown in Tab.~\ref{tab:low-lumi-scan-2},
from which one can get closer to the optimal energy point.
This process can be iterated with more times to get close enough to the best energy point.

\begin{table}[hbt!]
\centering
\begin{tabular}{l|c}
\hline
$\sqrt{s}$ (GeV) & $m_{top}$ precision (MeV) \\
\hline 
343 & 3.0 \\

342 & 3.5 \\

341 & 5.6 \\

\hline
\end{tabular}
\caption{\label{tab:low-lumi-scan}The precision of top quark mass measurement in a low-luminosity scan (1 fb$^{-1}$ per energy point). The centre-of-mass energies of \{340, 341, 342, 343, 344, 345\} are scanned. The scan points are ordered with the measured top quark mass precision. The precision shown in the table is scaled to an arbitrary number. The relative order matters. Only the first three are listed in the table.}
\end{table}

\begin{table}[hbt!]
\centering
\begin{tabular}{l|c}
\hline
$\sqrt{s}$ (GeV) & $m_{top}$ precision (MeV) \\
\hline 
342.75 & 2.9 \\

342.50 & 3.0 \\

343.00 & 3.0 \\

\hline
\end{tabular}
\caption{\label{tab:low-lumi-scan-2}The precision of top quark mass measurement in a low-luminosity scan (1 fb$^{-1}$ per energy point). The centre-of-mass energies of \{342.25, 342.5, 342.75, 343, 343.25, 343.5\} are scanned. The scan points are ordered with the measured top quark mass precision. The precision shown in the table is scaled to an arbitrary number. The relative order matters. Only the first three are listed in the table. }
\end{table}


With the CEPC setup, the ultimate statistical precision is calculated individually for top quark mass, width and $\alpha_S$ at their optimal energy points that are determined with a quick scan using low luminosity, respectively, assuming the total luminosity of 100 fb$^{-1}$ for each case. The statistical uncertainties of these measurements are listed in Tab.~\ref{tab:stats-only}. Relevant systematic uncertainties are discussed in the next section.

\begin{table}[hbt!]
\centering
\begin{tabular}{l|c|c|c}
\hline
Centre-of-mass energy & Precision of $m_{top}$ & Precision of $\Gamma_{top}$ & Precision of $\alpha_S$ \\
\hline 
342.75 GeV optimal for $m_{top}$ & 9 MeV & 343 MeV & 0.00041 \\
344.00 GeV optimal for $\Gamma_{top}$ & $>$ 50 MeV & 26 MeV & 0.00047  \\
343.25 GeV optimal for $\alpha_S$ & 15 MeV & 40 MeV & 0.00040 \\
\hline
\end{tabular}
\caption{\label{tab:stats-only} The statistical uncertainties of the measurements of top mass, width and $\alpha_S$ measured individually at their optimal energy points that can be determined with a quick scan using low luminosity.}
\end{table}

As already shown in Tab.~\ref{tab:stats-only}, the optimal energy point for $m_{top}$ cannot provide the best precision for $\Gamma_{top}$ and $\alpha_S$, as their optimal energies are all different. One can consider to use two energy points or more to reach a certain level of precision of $\Gamma_{top}$ and $\alpha_S$ while not to degrade too much the precision of $m_{top}$. Two energy points that have a total luminosity limited to 100 fb$^{-1}$ are discussed in the following.
Considering both $m_{top}$ and $\Gamma_{top}$, a various luminosity fractions between the two optimal energy points are tested, as shown in Tab.~\ref{tab:width-estimation}. Since the optimal energy point for $\Gamma_{top}$ is relative far away from the one for $m_{top}$, a small fraction like 20\% luminosity allocated for $\Gamma_{top}$ is already improving significantly its precision. A fraction even up to 50\% for $\Gamma_{top}$ still do not degrade too much the precision of $m_{top}$ but improves an order of magnitude in the precision of $\Gamma_{top}$. A similar test is performed for $m_{top}$ and $\alpha_S$, as shown in Tab.~\ref{tab:alphas-estimation}. The precision of $\alpha_S$ does not seem to change much when allocating luminosity to its optimal energy point, mostly due to the fact that the optimal energy points for $m_{top}$ and $\alpha_S$ are close.

\begin{table}[hbt!]
\centering
\begin{tabular}{l|c|c}
\hline
Lumi fractions for $\sqrt{s}=$[342.75,344.00] GeV& $m_{top}$ precision (MeV) & $\Gamma_{top}$ precision (MeV) \\
\hline 
[100\%, 0\%] & 9 & 343 \\

[80\%, 20\%] & 11 & 58 \\

[50\%, 50\%] & 13 & 36 \\

[20\%, 80\%] & 20 & 30 \\

[0\%, 100\%] & >50 & 26 \\

\hline
\end{tabular}
\caption{\label{tab:width-estimation}The precision of $m_{top}$ and $\Gamma_{top}$ with the total luminosity limited to 100 fb$^{-1}$. The first column gives the fractions of luminosity allocated to the optimal energy points for $m_{top}$ and $\Gamma_{top}$.}
\end{table}

\begin{table}[hbt!]
\centering
\begin{tabular}{l|c|c}
\hline
Lumi fractions for $\sqrt{s}=$[342.75,343.25] GeV & $m_{top}$ precision (MeV) & $\alpha_{S}$ precision \\
\hline 
[100\%, 0\%] & 9 & 0.00041 \\

[80\%, 20\%] & 10 & 0.00041 \\

[50\%, 50\%] & 11 & 0.00040 \\

[20\%, 80\%] & 13 & 0.00040 \\

[0\%, 100\%] & 15 & 0.00039 \\

\hline
\end{tabular}
\caption{\label{tab:alphas-estimation}The precision of $m_{top}$ and $\alpha_{S}$ with the total luminosity limited to 100 fb$^{-1}$. The first column gives the fractions of luminosity allocated to the optimal energy points for $m_{top}$ and $\alpha_{S}$.}
\end{table}


\section{Systematic uncertainties}
\label{sec:syst}

The threshold-scan method depends on the theoretical calculation of the cross-section curve as a function of the centre-of-mass energy. The uncertainty of the theoretical calculation is considered as 3\% based on conservative estimations from Ref.~\cite{Stahlhofen:2011iqb} and 1\% assumed to be achieved by the time of the experiments. This follows the same assumption as in Ref.~\cite{Seidel:2013sqa}.
The 1\% and 3\% uncertainty on the cross-section will lead to a measurement uncertainty of top quark mass 9 MeV and 26 MeV, which are in the same level of the statistical uncertainty and three times of that, respectively.

The background is considered to be subtracted cleanly from the observed data given the good signal-background separation in their shapes, such as the reconstructed top quark mass or a combined kinematic variable, and the statistical dominance of the signal events in the final fitting region. The background uncertainties are added to the likelihood function Eq.~\ref{eq:lh} as a nuisance parameter constrained by a Gaussian prior. The background efficiencies are taken from Ref.~\cite{Seidel:2013sqa}, and the cross-sections are calculated with Wizard V1.95~\cite{Kilian:2007gr,Moretti:2001zz}, as shown in Tab.\ref{tab:xsec_bkg}. Considering the background uncertainty as 1\% optimistically and 5\% conservatively, a measurement uncertainty of top quark mass of 4 MeV and 18 MeV is reached. From this, the background uncertainty is crucial. Measures like taking data below the threshold to constrain the background might need to be considered. 

\begin{table}[ !htpb ]
    \centering
    \begin{tabular}{c|cccc}\hline
$E_{\mathrm{cm}}$(GeV) & $qq$(fb) & $W^+W^-$(fb) & $ZW^+W^-$(fb) & $ZZ$(fb) \\\hline 
352 &$ 24149 \pm 69 $&$ 11628 \pm  4$&$    11.07 \pm  0.01$&$   703.5 \pm  0.3$\\
500 &$ 12136 \pm 46$&$  7708 \pm  3$&$    36.16 \pm  0.02$&$   447.9 \pm  0.2$\\
\hline 
    \end{tabular}
    \caption{Background cross-section near the top threshold and at 500 GeV.}
    \label{tab:xsec_bkg}
\end{table}

The variations in the beam energy could also lead to uncertainties on the top quark mass measurement. The beam energy uncertainty was reported at a level of 10$^{-4}$ in the operation of LEP~\cite{LEPEnergyWorkingGroup:2004mbb,OPAL:2004xxz} and the studies of ILC~\cite{Boogert:2009ir}, which already impacts the top quark measurement less than the statistical uncertainty as discussed in Ref.~\cite{Seidel:2013sqa}. In the CEPC scenario, the beam energy could vary 2.6 MeV as estimated from the accelerator team. This impacts the measurement of top quark mass maximally by 2 MeV, way below the statistical uncertainty. 


The other aspect from the beam is the uncertainty of the luminosity spectrum. Variations on the spread of the luminosity spectrum, i.e. the width $\sigma_{LS}$ in Eq.~\ref{eq:sigma_ls}, of 10\% and 20\% are considered. The corresponding uncertainties on the top quark mass measurement are 3 Mev and 5 MeV, respectively. These are quite different than the CLIC scenario in Ref.~\cite{Seidel:2013sqa} given the different controls of the luminosity spectrum in circular and linear colliders. Furthermore, the improvement of the top quark mass measurement by having a better luminosity spectrum, e.g. a smaller spread $\sigma_{LS}$, is evaluated. The reduction of the energy spread $\sigma_{LS}$ of 20\% and 50\% can lead to the statistical uncertainty of top quark mass measurement of 9.0 MeV and 8.4 MeV, with respect to our nominal statistical uncertainty of 9.1 MeV. It appears that a large improvement in luminosity spectrum does not bring much improvement in top quark mass measurement. This should be due to the excellent luminosity spectrum in circular colliders.

Taking into account all these uncertainties, the CEPC is expected to measure the top quark mass with a precision of 14 MeV and 34 MeV in the optimistic and conservative assumptions respectively, as shown in Tab.~\ref{tab:syst}

\begin{table}[hbt!]
\centering
\begin{tabular}{l|c|c}
\hline
Source & \multicolumn{2}{|c}{$m_{top}$ precision (MeV)} \\
& Optimistic & Conservative \\
\hline 
Statistics & 9 & 9 \\
\hline
Theory & 9 & 26 \\
Background & 4 & 18 \\
Beam energy & 2 & 2 \\
Luminosity spectrum & 3 & 5 \\
\hline
Total & 14 & 34 \\
\hline
\end{tabular}
\caption{\label{tab:syst}The expected statistical and systematical uncertainties of the top quark mass measurement in optimistic and conservative scenarios at CEPC.}
\end{table}

\section{Conclusions}
\label{sec:conclu}
We have studied the expected precision of the top quark mass, width and $\alpha_s$ in \ttbar production using an energy scan around the \ttbar threshold based on the CEPC scenario, assuming a total integrated luminosity of 100~\ifb. This study is performed with the package QQbar\_threshold (VERSION 2.2.0), including the effects of the initial state radiation and the luminosity spectrum.

At the threshold, the precision of the properties is highly dependent on the centre-of-mass energies in the scan. We use Fisher information to guide the choice of the scan points and find that there is an optimal energy point leading to the most precise measurement. With all the luminosity allocated to the single optimal energy point, the ultimate precision with the CEPC is 9 MeV for top quark mass, 26 MeV for top quark width and 0.00039 for $\alpha_s$ considering only the statistical uncertainty. The theory and the background uncertainties can have a large influence on the precision. Additional systematic uncertainties from the beam energy and the luminosity spectrum impact much less. The total uncertainty of top quark mass at CEPC can reach below 14 MeV optimistically and below 34 MeV conservatively at the optimal centre-of-mass energy.

In conclusion, the study shows that CEPC is capable of measuring the top quark mass with a precision below 34 MeV. The method requires a good understanding of theory and background estimations, and also requires a low-luminosity scan to identify the optimal energy point.

\acknowledgments

Z. Li, Y. Fang, G. Li, S. Xin, S. Wang, Y. Wang, Y. Zhang, H. Zhang are Z. Liang are supported in part by the IHEP innovative project on sciences and technologies under Project No. E2545AU210. X. Sun is supported in part by the National Science Foundation of China under Grants No. 12175006 and No. 12061141002, and by Peking University under startup Grant No. 7100603613.




\bibliographystyle{plain}
\bibliography{ourref}

\begin{thebibliography}{10}

\bibitem{CDF:2016vzt}
{Combination of CDF and D0 results on the mass of the top quark using up
  $9.7\:{\rm fb}^{-1}$ at the Tevatron}.
\newblock 8 2016.

\bibitem{CEPCStudyGroup:2018rmc}
{CEPC Conceptual Design Report: Volume 1 - Accelerator}.
\newblock 9 2018.

\bibitem{ATLAS:2018fwq}
Morad Aaboud et~al.
\newblock {Measurement of the top quark mass in the $t\bar{t}\rightarrow $
  lepton+jets channel from $\sqrt{s}=8$ TeV ATLAS data and combination with
  previous results}.
\newblock {\em Eur. Phys. J. C}, 79(4):290, 2019.

\bibitem{FCC:2018evy}
A.~Abada et~al.
\newblock {FCC-ee: The Lepton Collider}: {Future Circular Collider Conceptual
  Design Report Volume 2}.
\newblock {\em Eur. Phys. J. ST}, 228(2):261--623, 2019.

\bibitem{OPAL:2004xxz}
G.~Abbiendi et~al.
\newblock {Determination of the LEP beam energy using radiative fermion-pair
  events}.
\newblock {\em Phys. Lett. B}, 604:31--47, 2004.

\bibitem{CLICdp:2018esa}
H.~Abramowicz et~al.
\newblock {Top-Quark Physics at the CLIC Electron-Positron Linear Collider}.
\newblock {\em JHEP}, 11:003, 2019.

\bibitem{BESIII:2019}
M.N. Achasov, X.~Mo, N.Yu Muchnoi, Ivan Nikolaev, S.A. Privalov, and J.Y.
  Zhang.
\newblock Tau mass measurement at bes-iii.
\newblock {\em EPJ Web of Conferences}, 212:08005, 01 2019.

\bibitem{Aicheler:1500095}
M~Aicheler, P~Burrows, M~Draper, T~Garvey, P~Lebrun, K~Peach, N~Phinney,
  H~Schmickler, D~Schulte, and N~Toge.
\newblock {\em {A Multi-TeV Linear Collider Based on CLIC Technology: CLIC
  Conceptual Design Report}}.
\newblock CERN Yellow Reports: Monographs. CERN, Geneva, 2012.

\bibitem{LEPEnergyWorkingGroup:2004mbb}
R.~Assmann et~al.
\newblock {Calibration of centre-of-mass energies at LEP 2 for a precise
  measurement of the W boson mass}.
\newblock {\em Eur. Phys. J. C}, 39:253--292, 2005.

\bibitem{Beneke:2016kkb}
M.~Beneke, Y.~Kiyo, A.~Maier, and J.~Piclum.
\newblock {Near-threshold production of heavy quarks with
  $\tt{QQbar\_threshold}$}.
\newblock {\em Comput. Phys. Commun.}, 209:96--115, 2016.

\bibitem{Beneke:2013jia}
M.~Beneke, Y.~Kiyo, and K.~Schuller.
\newblock {Third-order correction to top-quark pair production near threshold
  I. Effective theory set-up and matching coefficients}.
\newblock 12 2013.

\bibitem{Beneke:2015kwa}
Martin Beneke, Yuichiro Kiyo, Peter Marquard, Alexander Penin, Jan Piclum, and
  Matthias Steinhauser.
\newblock {Next-to-Next-to-Next-to-Leading Order QCD Prediction for the Top
  Antitop $S$-Wave Pair Production Cross Section Near Threshold in $e^+e^-$
  Annihilation}.
\newblock {\em Phys. Rev. Lett.}, 115(19):192001, 2015.

\bibitem{Beneke:2017rdn}
Martin Beneke, Andreas Maier, Thomas Rauh, and Pedro Ruiz-Femenia.
\newblock {Non-resonant and electroweak NNLO correction to the $e^+ e^-$ top
  anti-top threshold}.
\newblock {\em JHEP}, 02:125, 2018.

\bibitem{Bigi:1986jk}
Ikaros I.~Y. Bigi, Yuri~L. Dokshitzer, Valery~A. Khoze, Johann~H. Kuhn, and
  Peter~M. Zerwas.
\newblock {Production and Decay Properties of Ultraheavy Quarks}.
\newblock {\em Phys. Lett. B}, 181:157--163, 1986.

\bibitem{Boogert:2009ir}
S.~Boogert et~al.
\newblock {Polarimeters and Energy Spectrometers for the ILC Beam Delivery
  System}.
\newblock {\em JINST}, 4:P10015, 2009.

\bibitem{CEPCStudyGroup:2018ghi}
Mingyi Dong et~al.
\newblock {CEPC Conceptual Design Report: Volume 2 - Physics \& Detector}.
\newblock 11 2018.

\bibitem{Fadin:1987wz}
Victor~S. Fadin and Valery~A. Khoze.
\newblock {Threshold Behavior of Heavy Top Production in e+ e- Collisions}.
\newblock {\em JETP Lett.}, 46:525--529, 1987.

\bibitem{Fadin:1988fn}
Victor~S. Fadin and Valery~A. Khoze.
\newblock {Production of a pair of heavy quarks in e+ e- annihilation in the
  threshold region}.
\newblock {\em Sov. J. Nucl. Phys.}, 48:309--313, 1988.

\bibitem{Horiguchi:2013wra}
Tomohiro Horiguchi, Akimasa Ishikawa, Taikan Suehara, Keisuke Fujii, Yukinari
  Sumino, Yuichiro Kiyo, and Hitoshi Yamamoto.
\newblock {Study of top quark pair production near threshold at the ILC}.
\newblock 10 2013.

\bibitem{CMS:2015lbj}
Vardan Khachatryan et~al.
\newblock {Measurement of the top quark mass using proton-proton data at
  ${\sqrt{(s)}}$ = 7 and 8 TeV}.
\newblock {\em Phys. Rev. D}, 93(7):072004, 2016.

\bibitem{Kilian:2007gr}
Wolfgang Kilian, Thorsten Ohl, and Jurgen Reuter.
\newblock {WHIZARD: Simulating Multi-Particle Processes at LHC and ILC}.
\newblock {\em Eur. Phys. J. C}, 71:1742, 2011.

\bibitem{Linssen:1425915}
Lucie Linssen, Akiya Miyamoto, Marcel Stanitzki, and Harry Weerts.
\newblock {\em {Physics and Detectors at CLIC: CLIC Conceptual Design Report}}.
\newblock CERN Yellow Reports: Monographs. CERN, Geneva, 2012.
\newblock Comments: 257 p, published as CERN Yellow Report CERN-2012-003.

\bibitem{Martinez:2002st}
Manel Martinez and Ramon Miquel.
\newblock {Multiparameter fits to the t anti-t threshold observables at a
  future e+ e- linear collider}.
\newblock {\em Eur. Phys. J. C}, 27:49--55, 2003.

\bibitem{Moretti:2001zz}
Mauro Moretti, Thorsten Ohl, and Jurgen Reuter.
\newblock {O'Mega: An Optimizing matrix element generator}.
\newblock pages 1981--2009, 2 2001.

\bibitem{Seidel:2013sqa}
Katja Seidel, Frank Simon, Michal Tesar, and Stephane Poss.
\newblock {Top quark mass measurements at and above threshold at CLIC}.
\newblock {\em Eur. Phys. J. C}, 73(8):2530, 2013.

\bibitem{LEP:2020w}
P.X. Shen, P.~Azzurri, C.X. Yu, M.~Boonekamp, C.~M. Kuo, P.~Z. Lai, B.~Li,
  G.~Li, H.~N. Li, Z.~J. Liang, B.~Liu, J.~M. Qian, and L.~S. Shi.
\newblock {Data-taking strategy for the precise measurement of the W boson mass
  with a threshold scan at circular electron positron colliders}.
\newblock {\em Eur. Phys. J. C}, 80:66, 2020.

\bibitem{Stahlhofen:2011iqb}
Maximilian Stahlhofen and Andre Hoang.
\newblock {NNLL top-antitop production at threshold}.
\newblock {\em PoS}, RADCOR2011:025, 2011.

\bibitem{Strassler:1990nw}
Matthew~J. Strassler and Michael~E. Peskin.
\newblock {The Heavy top quark threshold: QCD and the Higgs}.
\newblock {\em Phys. Rev. D}, 43:1500--1514, 1991.

\end{thebibliography}

\end{document}